\documentclass[12pt]{iopart}

\newcommand{\brkt}[2]{\langle#1|#2\rangle}
\usepackage{epsf,epsfig}

\begin{document}
\title[Optimal representations of quantum states by gaussians in phase space]{Optimal representations of quantum states by gaussians in phase space}

\author{Anatole Kenfack\dag\ Jan M Rost\dag\ and Alfredo M Ozorio de Almeida\ddag  \footnote[3]{To
whom correspondence should be addressed (kenfack@mpipks-dresden.mpg.de)}
}
\address{\dag\ Max Planck Institute for the Physics of Complex Systems
N\"othnitzer Strasse 38, 01187 Dresden, Germany}
\address{\ddag\ Centro Brasileiro de Pesquisas Fisicas, R Xavier Sigaud 150,
22290-180 Rio de Janeiro, Brazil}
\begin{abstract}
A two-step optimization is proposed to represent an arbitrary  quantum
state to a desired accuracy with the least number of gaussians in
phase space. The Husimi distribution of the quantum state provides the
information to determine the modulus of the weight for the gaussians.
Then, the phase information contained in the Wigner distribution is used
to obtain the full complex weights by considering the relative phases for
pairs of gaussians, the chords.
The method is exemplified with several excited states $n$ of the harmonic
and the Morse oscillators. A semiclassical interpretation of the number of
gaussians needed as a function of the quantum number $n$ is given.
The representation can also  be used to characterize Wigner and Husimi
distributions directly which do not originate in a quantum state.
\end{abstract}
\maketitle
\section{Introduction}
In many situations there is a need to represent a given quantum state or a
probability distribution of states in terms of a simple basis which can be
easily handled in further processing, e.g., for integration in matrix
elements. Gaussians form such a basis although it is overcomplete.
Gaussians have the additional advantage that they can adopt the form of
coherent states in phase space, thereby supplying a direct link to
classical mechanics. Hence, the formulation in terms of gaussians is also
very useful for a mixed quantum-classical or quasiclassical description of
a system. In particular the evolution of systems with many degrees of freedom
can often be described within classical molecular dynamics only. Here, one is
interested in obtaining optimal initial conditions by translating the initial
quantum state of the system into a suitable classical phase space distribution
which can be classically propagated \cite{schmidt02}. While all quantum
information on an amplitude level is lost in this approximation, we will
introduce a representation which can preserve basic quantum interferences by
construction even in the case of the purely classical propagation. Finally, it is of
general and basic interest to reconstruct amplitudes or the density matrix,
i.e., coherent information, from phase space distributions. This has been done
recently in terms of the moments in time of the Wigner distribution
\cite{johansen98}. Although there is no principal need to do so we will
concentrate on the Wigner distribution  due to the attention it has received
recently in different areas.  In several experiments the Wigner distribution
of microscopic systems has been measured \cite{lutterbach}. Beyond the quantum
context,  in the time-frequency phase space of signal processing, analysis of
the Wigner distribution \cite{cohen95} is an important tool \cite{cohen95}, as for optics
\cite{bartelt80}, speech analysis \cite{riley92} or the monitoring of machinery conditions \cite{meng91}.

The decomposition of a given state $|\Psi\rangle$ into coherent states can be
regarded as a particular form of phase space representation. Labeling
positions by $q$ or $Q$ and momenta by $p$ or $P$, we translate the ground
state of the harmonic oscillator by $X=(P,Q)$ to obtain the coherent state in position representation, i.e. its wavefunction is~\cite{CT,S}
\begin{eqnarray}
\brkt{q}{X}= \left(\frac{m\omega}{\pi \hbar}\right)^{1/4}\exp\left(-\frac{\omega}{2\hbar}(q-Q)^2+i\frac{P}{\hbar}\left(q-\frac{Q}{2}\right)\right).\label{e11}
\end{eqnarray}
For simplicity, we chose unit frequency ($\omega=1$) and mass ($m=1$) for the
harmonic oscillator without loss of generality. (Often the complex
variable $Z=(Q+iP)/\sqrt{\hbar}$ is used and the constant phase $-PQ/2\hbar$
is omitted). Even though the coherent state basis is overcomplete, the exact representation, \begin{eqnarray}
|\Psi\rangle=\frac{1}{\pi}\int dX |X\rangle \brkt{X}{\Psi},\label{e12}
\end{eqnarray}
is unique~\cite{CT,S}. However, such a decomposition into coherent states is
monstrously inefficient, as shown by the overlap of coherent states, 
\begin{eqnarray}
\brkt{X}{X'}&=&\exp\left(-\frac{(Q-Q')^2}{4\hbar}-\frac{(P-P')^2}{4 \hbar}
+\frac{i}{2\hbar}(QP'-Q'P)\right)\nonumber\\
&\equiv&\exp\left(-\frac{(X-X')^2}{4\hbar}+\frac{i}{2\hbar} X \wedge X'\right),\label{e13}
\end{eqnarray}
so that 
\begin{eqnarray}
|\brkt{X}{X'}|^2=\exp\left(-\frac{(X-X')^2}{2\hbar}\right)\label{e14}.
\end{eqnarray}
We have used the wedge product,  
\begin{eqnarray}
X\wedge X'=PQ'-QP'= ({\bf J} X)\cdot X',\label{e15}
\end{eqnarray}
to shorten the notation. (With the second equation we also define the symplectic matrix $\bf{J}$).
It follows that one can approximate any given state by a finite expansion 
\begin{eqnarray}
|\Psi\rangle\cong\sum_j a_j|X_j\rangle\label{e16}
\end{eqnarray}
with arbitrary accuracy. One possibility would be to place the coherent states
on a grid and then determine the complex coefficients $a_j$ by minimizing the
error, defined in some convenient way. Of course, it is much better to use
some features of $|\Psi\rangle$ as a preliminary guide to where the
$|X_j\rangle$ should be located.

An obvious necessary condition for a good fit is that the coherent state
intensity $|\brkt{X}{\Psi}|^2$ in $(\ref{e14})$ be well approximated by
\begin{eqnarray}
|\brkt{X}{\Psi}|^2&=&|\sum_j a_j \brkt{X}{X_j}|^2\nonumber\\
&=&\sum_j|a_j|^2|\brkt{X}{X_j}|^2+\sum_{j\neq k}a_j a_k^{*}\brkt{X}{X_j}
\brkt{X}{X_k}^{*}\label{e17}
\end{eqnarray}
for each phase space point $X$. But note that, if the diagonal terms in this
expression are only appreciable for $X_j$ very close to $X$, a nondiagonal
term decays as a product of gaussians centered on a pair of separate phase
space vectors.  Such a term is negligible  for all $X$ unless $X_j$ is very
close to $X_k$. Moreover, such nondiagonal  terms are not positive definite, which justifies our performing the preliminary fit
\begin{eqnarray}
\rho_{\Psi}(X)=\frac{1}{\pi}|\brkt{X}{\Psi}|^2
\approx \sum_j|a_j|^2\frac{1}{\pi}|\brkt{X}{X_j}|^2=\sum_j|a_j|^2 G_j(X),\label{e18}
\end{eqnarray}
where
\begin{eqnarray}
G_j(X)=\exp\left(-\frac{(X-X_j)^2}{2\hbar} \right).\label{e19}
\end{eqnarray}
The factor $1/\pi$ has here been introduced so that the left hand side becomes
the Husimi function \cite{H}, i.e., the Husimi representation of the state
$|\Psi\rangle$, also known as the $Q$-function in quantum optics
\cite{S}. Here, $\rho_\Psi$ is then decomposed into gaussians phase space
packets (gpps) which are Husimi representations of the coherent states,
specified by $(\ref{e14})$.

The Husimi function in its own right uniquely defines a quantum state, though
part of the quantum phase information is locked up into its delicate analytic
properties. These are hard to translate into other representations, but we
propose to sidestep this problem. Our method is to make a preliminary fit of
the Husimi function $(\ref{e18})$, which is real and positive definite, and
therefore much simpler than fitting the complex coherent states directly in $(\ref{e16})$. 
However we can now identify the coefficients of the Husimi fitting as being 
proportional to the square modulus of the coefficients of the superposition 
of coherent states $(\ref{e16})$. The phases of the coefficients are then
determined by relating the Husimi function to the coarse-graining of the
Wigner function, i.e. we bring in yet a third phase space representation of the state $|\Psi\rangle$.
It should be pointed out that, once the state $|\Psi\rangle$ has been fitted
by a finite number of coherent states, it can be viewed in any representation.
The special attention given here to the Wigner function is justified by the way
that it highlights quantum interferences. In fact such interferences are
sometimes mimicked  by extra gaussians in the context of numerical work
with classical densities~\cite{schmidt02}.
Thus, in the next section we present the Wigner function, its relation  
to the Husimi function and its specific form for an arbitrary
superposition of gaussians. In addition, we discuss and
establish the differences between the representation of quantum states such as
the wavefunction, the Wigner distribution and the Husimi distribution using
these gaussians. The Wigner distribution sets the stage for the delicate
procedure of fitting the phases, i.e., to determine the appropriate
interference fringes between the gaussians as detailed in section 3. In this
section we propose a method for determining the phase of the expansion
coefficients by Fourier analysing the Wigner function. Actually, this
resulting {\it chord function} or {\it characteristic function} (in quantum
optics~\cite{radmore97}) can be derived directly from the initial wave
function, without obtaining first the Wigner function. The Husimi function, on
the other hand, supplies a robust overall picture of the quantum states which can be used to place the gaussians. This will be
discussed in section 4 using explicit examples for illustration. Emphasis will
be put on general symmetry criteria for placing the gaussians and numerical
demonstrations based on Monte Carlo fitting of sampled wavefunctions. Finally
in section 5, we semiclassically interpret the results of section 4, namely
the increase of gaussians needed to represent quantum states $\psi_n$ of a
system with increasing quantum number $n$. A summary in section 6 concludes
the paper.

For simplicity, all our formulae are presented for systems with a single
degree of freedom, but they are easily generalized to high dimensional systems.
\section{Wigner functions and their relations}
The state $|\Psi\rangle$ is uniquely described by the real function of phase
space points $x=(p,q)$,
\begin{eqnarray}
W_{\Psi}(x)=\int \frac{dq'}{2\pi \hbar}\brkt{q+\frac{q'}{2}}{\Psi}\brkt{\Psi}{q-\frac{q'}{2}}
\exp\left(-\frac{i}{\hbar}pq'\right)\label{e21}
\end{eqnarray}
known as the Wigner function \cite{W}. Inversion of this symmetrized Fourier
transform leads back to the wavefunction, while its intensity results from the
projection
\begin{eqnarray}
|\brkt{q}{\Psi}|^2=\int dp W_{\Psi}(x).\label{e22}
\end{eqnarray}
Though the Wigner function can be negative for some points $x$, its projection
is always positive. Actually, this is just a particular case of the general
relation, 
\begin{eqnarray}
|\brkt{\Phi}{\Psi}|^2=2\pi\hbar\int dx W_{\Phi}(x) W_{\Psi}(x),\label{e23}
\end{eqnarray}
for arbitrary states $|\Phi\rangle$ and $|\Psi\rangle$. Thus, introducing the
Wigner function for the coherent state $|X\rangle$,
\begin{eqnarray}
W_{X}(x)=\frac{1}{\pi\hbar}\exp\left(-\frac{(x-X)^2}{\hbar} \right),\label{e24}
\end{eqnarray}
we obtain the Husimi representation of $|\Psi\rangle$ as
\begin{eqnarray}
\frac{1}{\pi}|\brkt{X}{\Psi}|^2&=&2\hbar\int dx W_{X}(x)W_{\Psi}(x)\nonumber\\
&=&\frac{2}{\pi}\int dx \exp\left(-\frac{(x-X)^2}{\hbar} \right)W_{\Psi}(x).\label{e25}
\end{eqnarray}
This shows that the positive definite Husimi function results from the
gaussian smoothing of the Wigner function.
Inserting the coherent state expansion $|\Psi\rangle$ given by $(\ref{e16})$
into the Wigner transform (\ref{e21}), we obtain the well known form of the corresponding Wigner function,
\begin{eqnarray}
W_{\Psi}(x)=\sum_{j}|a_j|^2W_{X_j}+2Re\left(\sum_{j\neq k}a_ja_k^{*}W_{jk}(x)\right),\label{e26}
\end{eqnarray}
where we define the crossed Wigner function (or Moyal bracket \cite{M}) 
\begin{eqnarray}
W_{jk}(x)&\equiv& \int \frac{dq'}{2\pi\hbar}
\brkt{q+\frac{q'}{2}}{X_j}\brkt{X_k}{q-\frac{q'}{2}}\exp\left(-\frac{i}{\hbar}pq'\right)\\
&=&\frac{1}{\sqrt{2}\pi\hbar}\exp\left(-\frac{(x-\bar{X}_{jk})^2}{\hbar}\right)
\exp\left(-\frac{i}{\hbar}\left(x \wedge \delta X_{jk} +
\frac{1}{2}X_k\wedge X_j\right) \right).\nonumber\label{e27}
\end{eqnarray}
Here we use the abbreviated relations $\bar{X}_{jk}=(X_k+X_j)/2$ and 
$\delta X_{jk}=X_j-X_k$, as well as the skew product $(\ref{e15})$.
Comparing $(\ref{e26})$ with $(\ref{e17})$ we find the same gaussian form for
the diagonal terms of the Wigner function as for the Husimi function, but the
offdiagonal terms are no longer negligible.
For the Wigner function, any pair of coherent states determines a third
gaussian centered halfway between their two centres, of comparable amplitude 
to the diagonal contributions of the individual gaussians. This interference
term  is only cancelled by the smoothing employed to obtain the Husimi
function, because of the phase oscillations. Their wavelengths in phase space
are $2\pi\hbar/|\delta X|$, so that the Husimi function is increasingly damped with growing separation of these gpps.
This is an excellent illustration of the way that the Husimi function
highlights the classical structure in a quantum state, while it hides quantum
phase information. This question will be further developed in section 4. Of
course the phase information is still contained in the small offdiagonal terms
in $(\ref{e17})$, because the Husimi function is fully quantum mechanical, but
this information is certainly more accessible in the Wigner representation. 

We can now understand the differences between using the position
representation, i.e. the wave function,  the Wigner representation or the
Husimi representation, to fit a given state by a set of coherent states. The
position representation has the advantage that there are no interference terms
between pairs of coherent states. However, if the state is highly oscillatory,
such as a highly excited eigenstate of a Hamiltonian, this needs to be fitted
by coherent states with large momenta $P$, which also have narrow
oscillations. Clearly the fitting procedure becomes very unstable with respect
to small errors in the initial function if large momenta are necessary. In
contrast, the diagonal terms in the expansion of the Wigner function are
smooth gaussians distributed over phase space, but again we obtain
increasingly narrow oscillations for the interferences around the midpoint of
each pair. Of course, the oscillations of the wave function and the Wigner
function are not unrelated, since the latter have the same wavelength as the
wave intensity in the case of a pair of gaussians located at ($\pm P,Q$).

Finally, the Husimi function washes out oscillations, the more effectively the
tighter they happen to be. We shall review in section 4 how the smooth phase
space distribution singles out the basic classical structure within the
quantum state, which can be stably fitted by real gpps. 

The cost is that only the location of the gaussians in phase space and the
square modulus of their coefficients are then determined. In the next section
we present a method for retrieving the phases of the coefficients by taking the
Fourier transform of the Wigner function.

\section{Fitting the phases}
Both, the direction and the wavelength 
of the fringes that modulate the interference peak halfway between two gpps 
at $X_j$ and $X_k$ in the Wigner representation, are entirely determined by
$\delta X_{jk}$. However, these fringes are translated by changing the phases
of the coefficients $a_j$ and $a_k$, which we wish to determine. Clearly, the
best way to recover these phases is to take the Fourier transform of the Wigner function:
\begin{eqnarray}
\tilde{W}_{\Psi}(\xi)&=&\int\frac{dpdq}{2\pi\hbar}W_{\Psi}(p,q)\exp\left(\frac{i}{\hbar}(p\xi_q-q\xi_p)\right)\nonumber\\
&=&\int\frac{dx}{2\pi\hbar}W_{\Psi}(x)\exp\left(\frac{i}{\hbar}(x\wedge \xi)\right).\label{e31}
\end{eqnarray}
It is also possible to express this directly from the wave function
\begin{eqnarray}
\tilde{W}_{\Psi}(\xi)=\int dq
\brkt{q+\frac{\xi_q}{2}}{\Psi}\brkt{\Psi}{q-\frac{\xi_q}{2}}\exp\big(-\frac{i}{\hbar}\xi_p q\big),\label{e32}
\end{eqnarray}
which resembles the Wigner transform $(\ref{e21})$.
This function is known as the Woodward ambiguity function in communications
theory \cite{Amb} or the characteristic function in quantum optics \cite{radmore97}. Usually the sign of $\xi_p$ is reversed in the definition of $\tilde {W}$, but here we follow \cite{Rep}, where it is simply called the chord function. A justification for this term appears immediately if we consider the case of a superposition of gaussians that is our concern :
\begin{eqnarray}
\tilde{W}_{\Psi}(\xi)=\sum_j|a_j|^2\tilde{W}_{jj}(\xi)+
\sum_{j\neq k}a_ja_k^{*}\left(\tilde{W}_{jk}(\xi)+\tilde{W}_{kj}(\xi)\right),\label{e33}
\end{eqnarray}
where 
\begin{eqnarray}
\tilde{W}_{jj}(\xi)=\frac{1}{2\pi\hbar}\exp\left(
-\frac{\xi^2}{\hbar}+\frac{i}{\hbar}X_j\wedge \xi\right)\label{e34}
\end{eqnarray}
and 
\begin{eqnarray}
\tilde{W}_{jk}(\xi)&=&\frac{1}{2\sqrt{2}\pi\hbar}\exp\left(
-\frac{(\xi-\delta X_{jk})^2}{4\hbar}\right)\nonumber\\
&\times & \exp\left(\frac{i}{\hbar}
\left((\xi-\delta X_{jk})\wedge\bar{X}_{jk} +\frac{X_k \wedge X_j}{2}\right)\right).\label{e35}
\end{eqnarray}
Thus all the diagonal terms collapse onto gaussians centered on $\xi=0$,
whereas each interference term is a gaussian centered on the chord joining
$X_j$ and $X_k$. (Both directions $\pm \delta X_{jk}$ are present, leading to
symmetric contributions in chord space. Indeed, we must have $\tilde{W}(-\xi)=
\tilde{W}^{*}(\xi)$).
Let us now suppose that a given chord $\delta X_{jk}$ is sufficiently far from
all the other chords, so that at $\xi=\delta X_{jk}$ the chord function is
dominated by the single nondiagonal contribution $\tilde{W}(\delta X_{jk})$,
i.e.
\begin{eqnarray}
\tilde{W}_{\Psi}(\delta
X_{jk})-\sum_l|a_l|^2\tilde{W}_{ll}(\delta X_{jk})&\approx&a_{j}a_{k}^{*}\tilde{W}_{jk}(\delta X_{jk})\nonumber\\
&=&\frac{a_ja_j^{*}}{2\sqrt{2}\pi\hbar}\exp\big(\frac{i}{\hbar}\frac{X_k \wedge X_j}{2}\big).\label{e36}
\end{eqnarray}
Then, defining 
\begin{eqnarray}
a_j=|a_j|\exp(i\theta_j)\label{e37}
\end{eqnarray}
and recalling that we have already fitted $|a_j|$ from the Husimi function, we
obtain $\theta_{jk}=\theta_j-\theta_k$ from
\begin{eqnarray}
\theta_{jk}-\frac{1}{2\hbar}X_j\wedge X_k=phase\Big(\tilde{W}_{\Psi}(\delta X_{jk})
-\sum_l|a_l|^2\tilde{W}_{ll}(\delta X_{jk})\Big).\label{e38}
\end{eqnarray}
Of course, the modulus of both sides of $(\ref{e36})$ should be approximately equal, which provides a check on the previous Husimi fitting. 
The diagonal sum that is substracted from $\tilde{W}_{\Psi}(\delta X_{jk})$ in
$(\ref{e36})$ decays exponentially with $\delta X_{jk}^2$, so it will only affect the phases of the smaller chords. 
It might appear to be only consistent with
our previous approximation of the Husimi function to neglect the diagonal sum in (\ref{e36}),
but this would perturb the phases of small chords.
\section{Placing the gaussians: examples}
So far, nothing has been said about how to place the gaussians in phase space
that are meant to approximate a given Husimi function. An obvious criterion is
that any relative maximum of this smooth distribution should also receive 
a coherent state. Further knowledge of the state should also be used. For
instance, if it is known to be the eigenstate of a given Hamiltonian
$\hat{H}$ with energy $E_n$, then the semiclassical considerations in the
following section allow the restriction of the gaussian centers to
$|H(X)-E_n|\le \hbar$.
Of course, the fitting procedure is only useful for cases in which $\hat{H}$
is unknown, or else $|\Psi\rangle$ may itself be a linear superposition of
states, such as result from a two slit experiment. Here we immediately feel the
advantage that the decomposition $(\ref{e16})$ is defined in terms of the state
itself, rather than the density operator $|\Psi\rangle\langle\Psi|$. Thus, the
linearity of $|\Psi\rangle=a_1|\Psi_1\rangle+a_2|\Psi_2\rangle$ allows us to
fit $|\Psi_1\rangle$ and  $|\Psi_2\rangle$ independently and then to
superpose the fitted state as
$|\psi\rangle=a_1|\psi_1\rangle+a_2|\psi_2\rangle$. 
A further simplication results from possible symmetries. For instance, for the
eigenstate of an even potential, one should place the gaussians symmetrically
at $(P_j,\pm Q_j)$ with $a_{j-}=\pm a_{j+}$, so as to guarantee even or odd
states. Again, this is a very particular case, but time reversal invariance is
much more common. If the state $|\Psi\rangle$ resulted from initial real
wave functions and the evolution proceeded through interactions that preserve
time reversal symmetries (in usual practice, if there are no magnetic fields)
then the final state can also be real. In this case, one must choose
symmetric pairs of gaussians at $(\pm P_j,Q_j)$ with equal real coefficients,
$a_{j-}=a_{j+}$, so as to ensure that $|\Psi\rangle$ is also real. This
halves the number of independent coherent states to be fitted and reduces 
the choice of phase to either zero, or $\pi$.
Any further knowledge should generally be used to reduce the randomness of the
positions of the coherent states to be fitted to $|\Psi\rangle$. The
semiclassical considerations in the following section may be a further
guideline. When this knowledge is exhausted, the best course is to optimize
random guesses by Monte Carlo fitting of arbitrary gaussians, within the given
constraints.

Specifically, in  order to measure the  quality of the fitting,  we define the
relative error
\begin{eqnarray}
\sigma(x,A,N)=\sum_{l=1}^M\left ( \rho_{\psi}(x_l)-\sum_{k=1}^N
A_kG_k(x_l)\right )^2 \Big /\sum_{l=1}^M\rho_{\psi}^2(x_l)\label{e41}
\end{eqnarray}
as the mean square deviation between the known Husimi 
function $\rho_\psi(x)$ of a quantum state $|\psi\rangle$ and its fitting to
the superposition of gpps defined in (\ref{e19}).
This error $\sigma $ has to be minimized for a set of $M$ grid points
$x_l=(p_l,q_l)$. To fit this superposition of gpps as closely as
possible to $\rho_\psi(x)$, we proceed as follows: (i) the centers
$x_k=(p_k,q_k)_{k=1,..,N}$ of the coherent states are obtained by Monte-Carlo
sampling the region where the distribution is significant; that is the region
where the distribution exceeds a certain threshold $\delta$; (ii) the centers
must not be too close, this means a minimum distance $\Delta$ between centers
is fixed a priori, thereby avoiding overlaps and reducing the total number $N$
of gpps used for the representation; (iii) starting with $N=1$ we increase N
by one and repeat the process until the desired accuracy is reached. Since
only the coefficients $A_k$ are fitted, minimizing the error $\sigma$ reduces
to a problem of linear optimization which is equivalent to solving the
real matrix equation 
\begin{eqnarray}
GA=\rho_\psi\label{e42}
\end{eqnarray} 
where the phase space points $x_l=(p_l,q_l)_{l=1,..,N}$ that define the matrix
elements $G_{lk}=G_k(x_l)=\exp\Big(-\frac{(x_l-x_k)^2}{2\hbar}\Big)$ are randomly
selected. Here,  $A=(A_1,A_2,...,A_N)$  contains the coefficients to be
determined and $\rho_\psi=(\rho_1,\rho_2,...,\rho_N)$ those of the known
Husimi function. In general, the matrix $G$ is expected to be sparse and a
variety of special algorithms can be used to solve (\ref{e42}) efficiently~\cite{k1}.
In principle one may also
consider to optimize this fitting with respect to the widths of the
gpps; but since the Husimi function is always smooth and positive definite, it is convenient to keep them frozen.
In what follows, we have tested our method on the fitting of the eigenstates of
the harmonic oscillator (HO) and that of the Morse oscillator (MO) supporting
18 bound states~\cite{k2}. We have not used the knowledge that these states have time reversal symmetry
in the way mentioned earlier, so we should obtain results of comparable quality
even for states with complex wave functions.
\begin{figure}
\begin{center}
\epsfbox{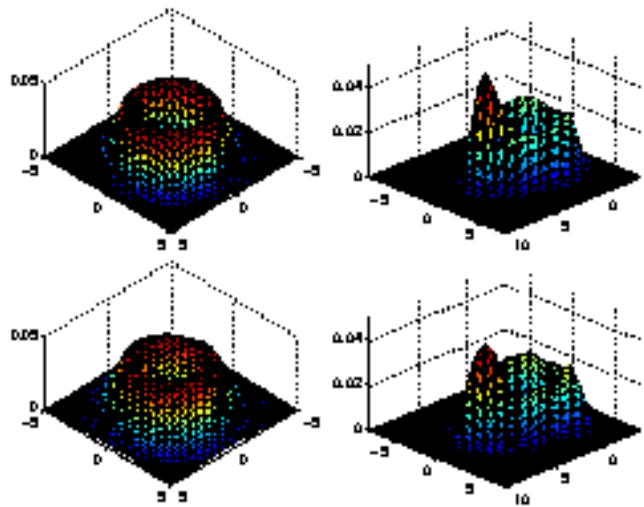}
\end{center}
\caption{\label{Figure 1}Husimi distributions of the 3rd excited state of the
harmonic oscillator (HO) and for the 5th excited state of the Morse oscillator
(MO); (top : original, bottom: fit). The fitting of HO (left) is reproduced
with $N=7$  ($\delta=0.035$, $\Delta=0.5$) and that of MO (right) with $N=14$ ($\delta=0.02$, $\Delta=1.3$).}
\end{figure}
Figure 1 shows, on the top, the known Husimi function for  the 3rd excited
state of HO (left) and for the 5th excited state of MO (right), with
parameters $\omega=m=\hbar=1$. On the bottom the corresponding fitted
distributions are shown. We found that $N=7$ is sufficient to reproduce quite
well the Husimi function of HO whereas 14 coherent states (gpps) are needed in
the MO example, to achieve a global relative error of $\sigma \le 0.01$. The
respective values for the threshold $\delta$ and the minimum distance $\Delta$
are given in the caption of figure 1. This preliminary fit on the Husimi function provides us with the total number $N$ of gpps, their location in phase
space $(p_k,q_k)$ and the expansion coefficients $A_k$. These coefficients are
related to those of the coherent states $a_k$ and their phases $\theta_k$ by 
\begin{eqnarray}
a_k=\sqrt{2\pi\hbar A_k}\exp(i\theta_k).\label{e43}
\end{eqnarray} 
\begin{figure}
\begin{center}
\epsfbox{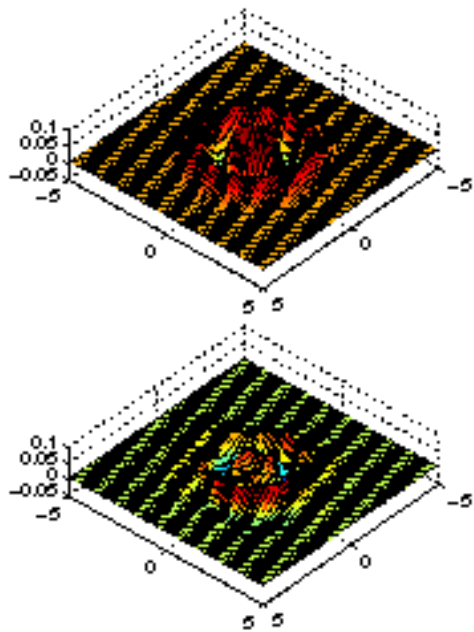}
\end{center}
\caption{\label{Figure 2}Wigner distributions of the 3rd excited state of HO
(top) and its corresponding fit (bottom) with $N=7$ as in figure 1.}
\end{figure}
\begin{figure}
\begin{center}
\epsfbox{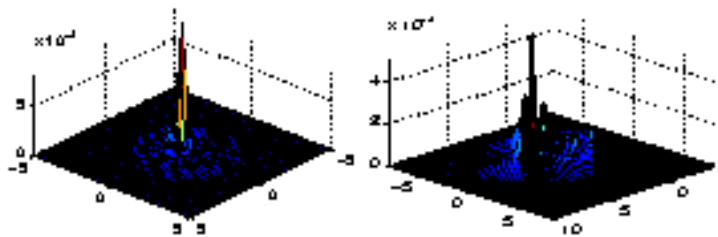}
\end{center}
\caption{\label{Figure 3}Deviations of the Monte Carlo Wigner sampled of the
3rd excited state of HO (left) and of the 5th excited state of MO (right).}
\end{figure}
\begin{figure}
\begin{center}
\epsfbox{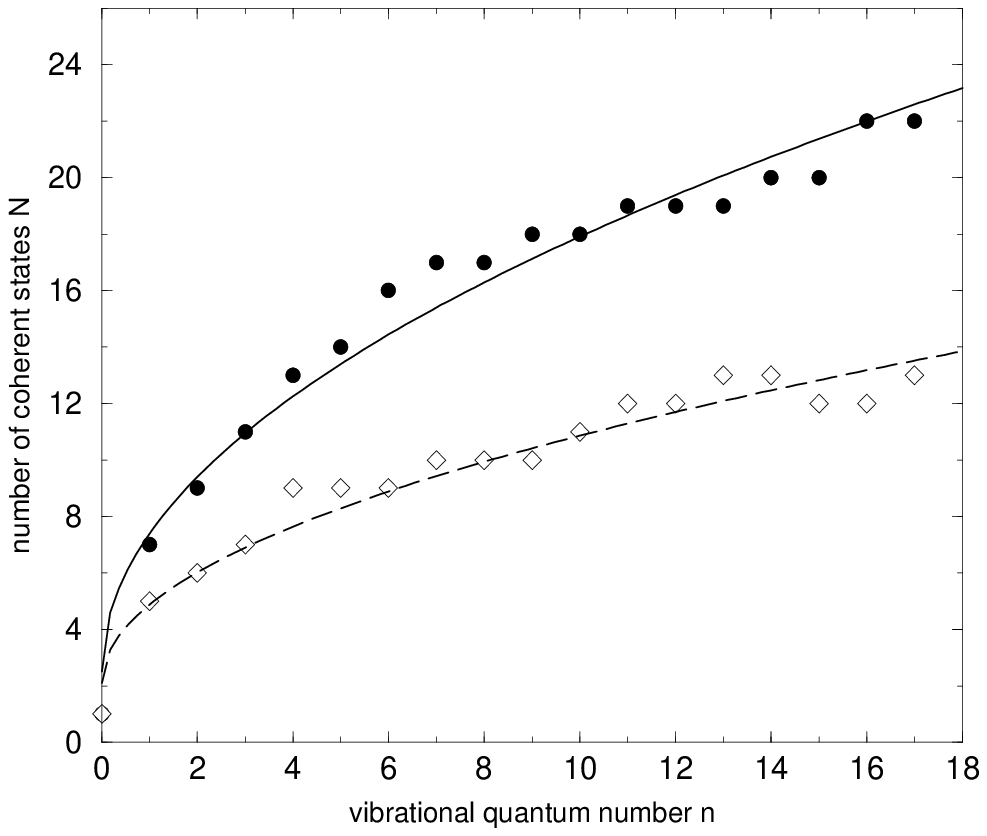}
\end{center}
\caption{\label{Figure 4}Number of coherent states $N$ used for the fitting
versus the vibrational quantum number $n$ for both, harmonic
oscillator (\opendiamond, dashed line: fit with $N=2.77\sqrt{n}+2.10$) and Morse
oscillators (\fullcircle, solid line: fit with $N=4.87\sqrt{n}+2.51$).}
\end{figure}
Up to this point, the phases $\theta_k$ are still unknown.
We only need to determine the phases differences $\theta_{jk}$ in (\ref{e26})
which can be easily extracted from (\ref{e38}). This says that, for $N$ gpps,
$N(N-1)/2$ chords ($\theta_{jk}$) are required to reproduce the Wigner
function. These $N(N-1)/2$ pairs of gaussians (chords) are symmetrically
spaced about $\xi=0$, so that, being more dense, they overlap more than the
original set of $N$ gaussians in phase space. Figure 2 shows on the top the
Wigner function of the 3rd excited state of the HO together with its
corresponding fit obtained with $7$ gpps. The present fitting reproduces the
Wigner function everywhere in phase space, even for negative regions which are
signatures of quantum interference. This is clearly illustrated in figure 3
where the deviations of the Monte Carlo sampled Wigner function, less than
$10^{-4}$ everywhere in phase space, are depicted for both, the HO (left,
$n=3$) and the MO (right, $n=5$). The global relative error $\sigma$ is
subsequently found not to exceed $0.01$. Moreover, we have computed the number
$N$ of gpps needed to represent the Wigner functions of the eigenstates of our
test systems, as can be seen in figure 4. One sees that $N$ grows
proportionally to $\sqrt{n}$ as the vibrational quantum number $n$  increases.
In the following section, we provide a semiclassical justification of this
result by means of the WKB quantization. 

\section{Semiclassical approach}
A semiclassical state $|\Psi\rangle_{SC}$ is supported by a curve in phase
space, in the case of one degree of freedom, or in general by a surface with
half the  phase space dimension \cite{VV,book}. If the curve or surface is
closed, it must be Bohr-quantized  and it will be symmetric about $p=0$ in the
case of time reversal symmetry. In the case of the position representation,
for each branch of the curve (surface), $p_j(q)$, one defines the action
\begin{eqnarray}
{\cal S}_j(q)=\int_{q_0}^q p_j dq,\label{e51}
\end{eqnarray}
leading to the generalized WKB wavefunction,
\begin{eqnarray}
\langle q|\Psi\rangle_{SC}= \sum_j a_j(q) \exp\left({i\over \hbar} {\cal S}_j(q)\right).\label{e52}
\end{eqnarray}
The amplitudes $a_j(q)$ can also be expressed in terms of the actions, 
but the important point is that they are purely classical so that the only
$\hbar$-dependence occurs in the exponential. Thus locally, for any small
range of positions, the semiclassical wave function reduces to a superposition 
of plane waves characterized by the wave-vectors $p_j(q)/ \hbar$. 
The semiclassical Wigner function is also defined in terms of an action ${\cal
S}_j(x)$ with respect to the classical curve (surface). But instead of the
area between the curve  and the $q$-axis, we are now concerned with the area
sandwiched between the curve and one of its chords. The latter, $\xi_k(x)$, is
selected by the property that  it is centred on the point $x$, as shown in figure 5. Thus, for one degree of freedom \cite{Trans},
\begin{eqnarray}
W_{\Psi}(x)_{SC}=\sum_k A_k(x) \cos\left({{\cal S}_k(x)\over \hbar}-{\pi\over 4}\right),
\label{e53}
\end{eqnarray} 
with straightforward generalizations \cite{OAH}. The important point is that \cite{Rep}
\begin{eqnarray}
\frac{\partial{\cal S}_k}{\partial x}=\bf{J} \xi_k,
\label{e54}
\end{eqnarray}
where we use the symplectic matrix $\bf J$ defined in (\ref{e15}).
Thus the semiclassical state is again represented by a superposition of waves,
\begin{eqnarray}
W_\Psi (x)_{SC}\approx \sum_k A_k(x) \cos \left(\frac{\delta x \wedge {\xi}_k(x)}{\hbar}-\frac{\pi}{4} \right),\label{e55}
\end{eqnarray}
\begin{figure}
\begin{center}
\epsfbox{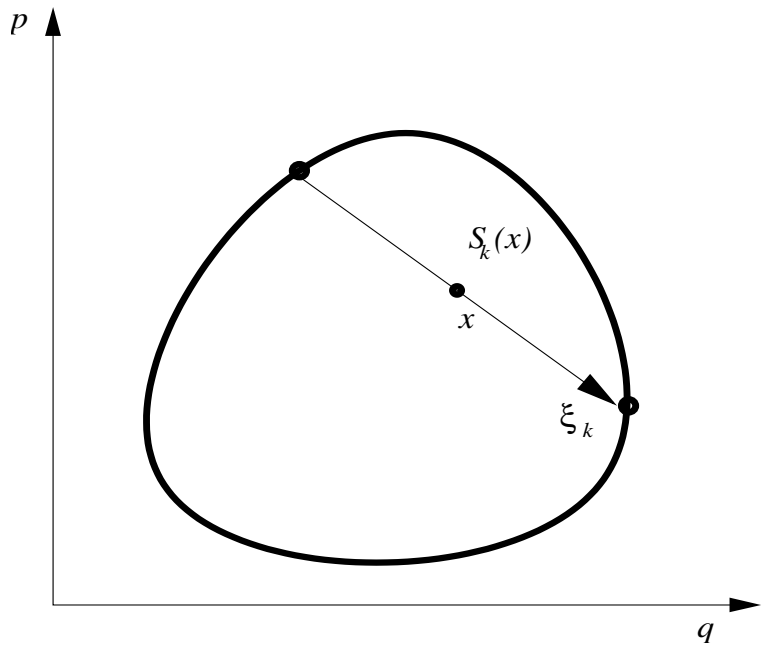}
\end{center}
\caption{\label{Fugure 5}Geometrical illustration of the chord $\xi_k(x)$ in
phase space $x(p,q)$.}
\end{figure}
but now these are {\it phase space waves}. Comparison of this expression with
(\ref{e36}) immediately reveals that these waves have exactly the same wave vector as a pair of gaussians at the tip of each chord $\xi_k(x)$.
It follows that an important feature of a semiclassical state is
automatically reproduced  by fitting it with gaussians precisely placed on the corresponding classical curve (surface).
Notice that, in fact, this is a major a priori obstacle to performing the
fitting,  because the oscillations near the midpoint of a long chord are very fine for $\hbar<<{\cal S}_k$.
One would need very narrow gaussians indeed to fit these phase space waves
directly, rather than having them arise naturally as interferences. All that
is left to determine the phase: ${\cal S}_k/\hbar-\pi/4$.
How can we single out the region in which to place the gpps, if we have no a
priori knowledge of the classical structures correponding to a quantum state?
The obvious course is to smooth the Wigner function with a Gaussian window. In other words, we initially fit the Husimi function.
Recalling that the gpps have linear width of order $\sqrt\hbar$, while the
wavelength of the Wigner oscillations are $\hbar/|\xi_k|$, it follows that
this gaussian window erases effectively a semiclassical Wigner function,
except in the limit of very small chords. This is a simple explanation of the well known fact that only in the neighbourhood of the classical curve (surface) 
itself is the semiclassical Husimi function appreciable. In the simple case of
one degree of freedom,  the Husimi function is concentrated near the energy
shell. Thus, by fitting gpps to a Husimi function, one automatically samples
its relevant classical mainifold, if it happens to have a (possibly unknown) semiclassical structure.
A basic assumption in the above argument is that the gpps along the curve
(surface) are neither too crowded, to avoid confusing superpositions, nor too sparsely spaced, lest gaps should arise in the fitting. It is thus easy to estimate the growth of the number of coherent states required to fit the eigenstate $|n\rangle$ with the degree of excitation.
Since the linear width of the gaussian scales as $\sqrt{\hbar}$, the phase
space area of the curve grows as $n\hbar$ and hence its length grows as
$\sqrt{n\hbar}$, the number of equispaced gaussians needed to cover the curve
grows as $\sqrt n$. Of course, we have assumed here that the shape  of the curve does not change with $n$, as for the harmonic oscillator. If the eigencurve elongates for higher excitation, the number of gpps necessary for a good fit will grow as $n^\alpha$, with $1/2\leq\alpha\leq 1$.
We can also estimate the growth in time of the number of gpps required to fit
an evolving semiclassical state with arbitrary precision. According to the
theory of van Vleck \cite{VV}, it is sufficient to evolve the curve (surface) classically and then to reconstruct the wave function.
This principle can also be applied to Wigner functions, according to Berry and Balazs~\cite{BB}.
If the driving Hamiltonian is chaotic, then the curve will stretch at a rate
depending on the Lyapunov exponent, $\lambda$. Clearly, this also determines
the initial rate of growth of the number of gpps needed for an adequate
fit. Ultimately, when the curve covers densely all the available phase space
(the energy shell of the driving Hamiltonian), a steady state will be reached
where the number of gpps saturates.

\section{Conclusions}
Optimal fitting of quantum states by phase space gaussians is achieved by
first fitting the Husimi function and then determining the quantum phases from
the chord function, i.e. the Fourier transform of the Wigner function. It is
clear from the discussion in the introduction that the density of gaussians in
phase space must be finely adjusted: If ther are too few, essential features
of the state will be missed, whereas an excessive number of gaussians would
introduce interference terms in the Husimi function itself, which could only
be accomodated by a much more complex variation of this method. The numerical
examples in section 4 indicate clearly that , as well as achieving excellent overall accuracy, all essential qualitative features of the states are captured
by this method, using a basis of gaussians that grows more slowly than the
excitation number of the fitted states.  
This last result and further insight into the fitting procedure, follows from
the semiclassical analysis in the previous section. This can be generalized
readily to states of quantum systems with higher degrees of freedom if these
are eigenstates of integrable systems (i. e. if they are supported by a Lagrangean surface in phase space; see e. g. \cite{book}).
Of course, more gaussians will then be needed for the fitting, but the preliminary fit of the Husimi function should still provide optimal results. The potential for this method to deal with the eigenstates of chaotic systems is even more interesting.
Though the {\it chord} and {\it centre} description of the Wigner function
still applies to mixtures of eigenstates over narrow energy windows, no
classical theory accounts for individual eigenstates at present. Therefore, it
will be extremely useful to describe these as interfering superpositions of
gaussians placed near the energy shell. Of course, the size of this basis
would diverge at the classical limit as $\hbar \rightarrow 0$, but manageable approximations should be attainable for finite excitations.

\ack
AK gratefully acknowledges the financial support by Alexander von Humboldt
(AvH) Foundation/Bonn-Germany, under the grant of Research fellowship
No.IV.4-KAM 1068533 STP. AMOA thanks the MPIPKS-Dresden for a Martin
Gutzwiller Fellowship, during which this work was initiated, and acknowledges
support from CNPq and Pronex in Brazil.

\end{document}